\documentclass[12pt]{iopart}
\usepackage[T1]{fontenc}

\usepackage{graphicx}
\usepackage{amssymb}
\usepackage{color}
\usepackage{psfrag}
\usepackage{epstopdf}
\usepackage{dsfont}

\usepackage{soul}

\newcommand{\bra}[1] {\langle #1 |}
\newcommand{\ket}[1] {| #1 \rangle}

\newcommand{\one}{\leavevmode\hbox{\small1\normalsize\kern-.33em1}}

\newcommand{\integlim}[3]{\ensuremath{\int_{#2}^{#3} \!\!\! \mathrm{d}#1 \,}}

\usepackage{iopams}  
\begin{document}

\title{Generation of Mechanical Interference Fringes by Multi-Photon Counting}

\author{M Ringbauer$^{1,2,3}$, T J Weinhold$^{1,2}$, L A Howard$^{1,2}$, A G White$^{1,2}$ \& M R Vanner$^{1,2,4,5}$}
\address{$^1$ Centre for Engineered Quantum Systems, School of Mathematics and Physics, University of Queensland, Brisbane, QLD 4072, Australia, \\
$^2$ Centre for Quantum Computer and Communication Technology, School of Mathematics and Physics, University of Queensland, Brisbane, QLD 4072, Australia \\
$^3$ Institute of Photonics and Quantum Sciences, School of Engineering and Physical Sciences, Heriot-Watt University, Edinburgh EH14 4AS, United Kingdom, \\
$^4$ Clarendon Laboratory, Department of Physics, University of Oxford, OX1 3PU, United Kingdom, \\
$^5$ QOLS, Blackett Laboratory, Imperial College London, London SW7 2BW, United Kingdom}
\ead{m.vanner@imperial.ac.uk}

\begin{abstract}
Exploring the quantum behaviour of macroscopic objects provides an intriguing avenue to study the foundations of physics and to develop a suite of quantum-enhanced technologies. One prominent path of study is provided by quantum optomechanics which utilizes the tools of quantum optics to control the motion of macroscopic mechanical resonators. Despite excellent recent progress, the preparation of mechanical quantum superposition states remains outstanding due to weak coupling and thermal decoherence. Here we present a novel optomechanical scheme that significantly relaxes these requirements allowing the preparation of quantum superposition states of motion of a mechanical resonator by exploiting the nonlinearity of multi-photon quantum measurements. Our method is capable of generating non-classical mechanical states without the need for strong single-photon coupling, is resilient against optical loss, and offers more favourable scaling against initial mechanical thermal occupation than existing schemes. Moreover, our approach allows the generation of larger superposition states by projecting the optical field onto NOON states. We experimentally demonstrate this multi-photon-counting technique on a mechanical thermal state in the classical limit and observe interference fringes in the mechanical position distribution that show phase super-resolution. This opens a feasible route to explore and exploit quantum phenomena at a macroscopic scale.
\end{abstract}



\submitto{\NJP}

\maketitle

\section{Introduction}

Generating and studying quantum superposition states of macroscopic objects is an important goal in many areas of experimental quantum science. Studying such states enables the limits of applicability of quantum mechanics to be probed and the exotic properties of quantum physics to be harnessed for the development of new technologies. Some of the earliest evidence for quantum phenomena with massive systems was provided by electron diffraction experiments~\cite{Davisson1927}. Then, through the efforts of the last nine decades, quantum-matter-wave behaviour has now been observed for neutrons~\cite{Rauch1974}, trapped ion systems~\cite{Monroe1996}, ultracold atoms~\cite{Shin2004}, and even molecules comprising many hundreds of atoms~\cite{Eibenberger2013}. A promising route to explore quantum behaviour on an even more macroscopic scale is provided by quantum optomechanics~\cite{RMP2014} where a mechanical resonator interacts with an optical field via radiation pressure. This versatile quantum-optical platform enables tests of fundamental physics~\cite{BJK99, Marshall03, Pikovski12}, the development of microwave-to-optical interfaces~\cite{Andrews14}, high-precision weak-force sensors~\cite{Rugar04}, and there is currently considerable interest towards generating mechanical superposition states, see e.g. Refs~\cite{Vanner11, Sekatski14, Ge15, Liao16, Hoff16}. Recently, impressive progress has been made using both opto- and electro-mechanical systems with examples including single-phonon-level operations~\cite{OConnell10, Lee2012, Vanner13, Riedinger2016, Hong2017}, quantum coherent coupling~\cite{Verhagen12}, mechanically-induced squeezing of light~\cite{Safavi13, Purdy13}, field-mechanics entanglement~\cite{Palomaki13}, and entanglement of two mechanical resonators~\cite{Riedinger2017, Ockeloen2017}. Experimental efforts continue in a diverse set of directions, however, progress is hindered by three main factors: weak single-photon coupling, sensitivity to optical loss, and mechanical decoherence. The approach we introduce here significantly relaxes these challenges by utilizing the measurement-induced nonlinearity of multi-photon counting. Measurement-based nonlinearities have been immensely successful in shaping the field of quantum photonics~\cite{KLM01} and single-photon counting has recently began to be explored in quantum optomechanics for single phonon addition/subtraction operations~\cite{Lee2012, Vanner13, Riedinger2016}. This work utilises photon counting in a qualitatively different way to what has been explored thus far and is the first to explore multi-photon counting operations for mechanical state preparation. Specifically, these photon-counting measurements provide an event-ready signal indicating that a mechanical resonator has interacted with an optical NOON state, thus generating a mechanical superposition state via the coherent $N$-photon quantum radiation-pressure interaction. Unlike many approaches throughout quantum optics, this scheme does not require strong single-photon coupling and is resilient against optical loss. Furthermore, the protocol is employed on a short time-scale and can generate non-classicality with a more favourable scaling against initial thermal occupation than previously reported. This experimental technique provides a powerful platform to empirically explore open-quantum-system dynamics, test potential collapse models of the wavefunction~\cite{GRW86,Diosi89,Penrose96}, and enable the development of quantum-enhanced weak force sensors.

To understand the radiation pressure interaction between light and a mechanical resonator, consider a single prompt reflection by a single photon. The reflection imparts a momentum---inversely proportional to the photon's wavelength---onto the resonator, and concurrently the optical field acquires a phase-shift---proportional to the mechanical displacement. The momentum transferred to the mechanical resonator is typically very small compared to its quantum noise due to the weakness of the radiation-pressure interaction. In order to enhance the strength of this interaction, experimental efforts often employ an optical cavity to increase the number of reflections~\cite{RMP2014}. Utilising such cavity enhancement, the seminal works of Bose, Jacobs, and Knight~\cite{BJK99}; and Marshall \etal~\cite{Marshall03}, proposed using a superposition of the optical vacuum and a single photon to generate optomechanical entanglement with the motion of a mechanically oscillating mirror forming part of a Fabry-Perot cavity, which opened an avenue to explore mechanical decoherence mechanisms.

Our method takes a very different tack to these two proposals and provides a feasible path to generate large-scale single-mode mechanical superposition states. In contrast to Refs~\cite{BJK99, Marshall03}, where the generation of optomechanical entangled states is proposed, our scheme allows mechanical non-classicality, such as Wigner negativity, to be probed and utilised directly. Specifically, we use an optomechanical interaction for a time much shorter than the mechanical period~\cite{Vanner2011} with a weak optical coherent state and then project the reflected field, via photon counting, onto a superposition of zero and one photon to generate the superposition. In this case, the mechanical resonator is subject to a quantum superposition of the identity operation (no photon present) and a displacement operation (single photon present), thus generating a mechanical Schr\"{o}dinger-cat state. It is important for the interaction to be shorter than the mechanical period so that the mechanical free evolution does not `smear' the superposition along the position axis. This is easily achieved in a cavity optomechanics context utilising pulses and operating outside the resolved sideband regime, i.e.\ our scheme requires that the cavity decay rate $\kappa$ be much larger than the mechanical frequency. A further advantage of our scheme is that no non-classical optical input states, such as single photons, are required, which can be challenging to prepare to match high-finesse optomechanical cavities. Such a single-mode mechanical quantum superposition state, with components separated in the momentum quadrature, will show interference fringes in the position probability distribution with a frequency inversely proportional to the superposition separation. After the photon counters give the event-ready signal for mechanical state preparation, an independent readout beam is used to verify and characterise the mechanical state. We would like to highlight that this quantum state preparation process is non-classical and generates quantum states of motion independent of the coupling strength and for finite initial thermal occupation.

To make larger mechanical superposition states one must engineer an interaction with an effective optical superposition that has a larger variance in the photon number. This variance is maximized for optical NOON states~\cite{Boto2000, Walther2004, Mitchell2004}, which are well known for quantum-enhanced metrology applications, however,  very challenging to prepare experimentally. In our scheme we instead propose to generate large mechanical superposition states by \emph{projection} onto optical NOON-states, which can be much more easily implemented using coherent state inputs, multi-port interferometry, and multi-photon coincidence measurements~\cite{Resch07}. Unlike \emph{preparation} of large NOON states, \emph{projection} onto these states is experimentally simple. We implement this process experimentally and observe the first mechanical interference fringe pattern and the predicted phase-superresolution for a two-photon coincidence measurement. In this proof-of-concept experiment we observe a high-visibility fringe pattern in the mechanical position distribution in a classical thermal regime, where no significant non-classicality could be observed. Generating and observing superposition states of a massive mechanical resonator remains an outstanding challenge. Our measurement-based scheme offers a promising route to achieving this highly sought goal and can be readily applied to a number of optomechanical systems beyond the system used here.

\section{Setup and Scheme}

\subsection{Mechanical state preparation via multi-photon detection}
\begin{figure}[t!]
\centering
\includegraphics[width=\linewidth]{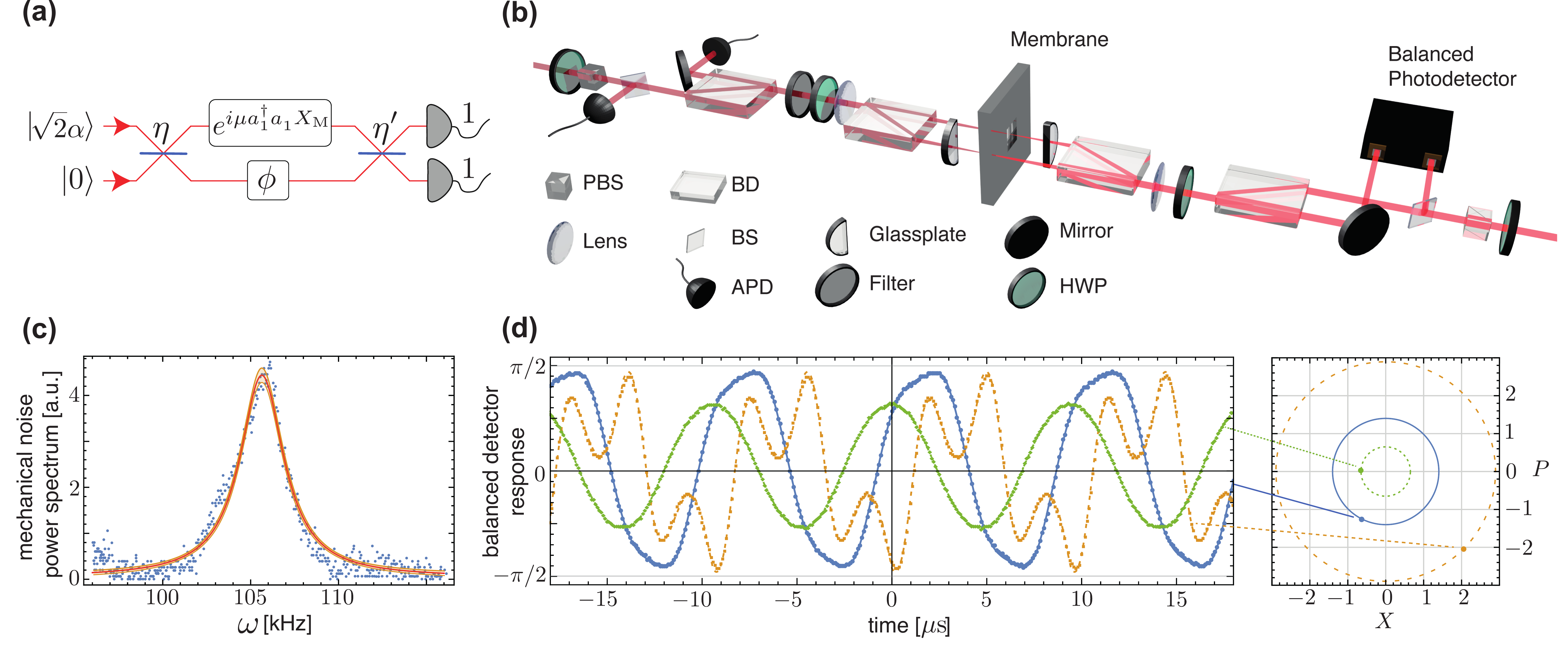}
\vspace{-1em}
\caption{\textbf{Experimental scheme.}
(\textbf{a}) A mechanical resonator interacts with a weak optical coherent state inside an interferometer with static phase-shift $\phi$ formed by two beam splitters with reflectivities $\eta$ and $\eta'$, respectively. The optomechanical radiation-pressure interaction is described via the unitary operation $e^{i\mu a_1^\dagger a_1 X_\textsc{m}}$, where $a_1$ describes the optical field operator, and $X_\textsc{m}$ describes the mechanical position. Mechanical interference fringes are generated via photon counting on the two interferometer outputs, which projects the field inside the interferometer onto a path-entangled photon number state.
(\textbf{b}) In our experiment, the two interferometer paths in A are represented by orthogonal polarizations of a weak coherent-state,  and a half-wave plate (HWP) acts as a tunable beam splitter. For the interaction with the mechanical resonator---a SiN membrane---the polarization modes are split into distinct optical paths using a calcite beam displacer (BD). One of the beams reflects off the mechanical membrane, while the other one reflects off the adjacent frame and acquires a static phase shift $\phi$, controlled by the yaw degree of freedom of the BD. The modes are then interfered, separated, and detected by two single-photon detectors (APD). A position measurement of the membrane is performed from the other side of the membrane using a similar setup and an independent readout beam. Lenses focus the preparation and readout beams to a spot size of ${\sim}50~\mu$m on the membrane and glass plates are used to compensate for shifts in the foci due to the birefringence in the BDs.
(\textbf{c}) Noise power spectrum of the fundamental vibrational drum mode of the mechanical resonator with resonance frequency of $\omega_{\textsc{m}}/2\pi = 105.64$~kHz.
(\textbf{d}) Balanced-detector time-traces used to measure the mechanical state for three example phase space points (shown on the right) with displacements of $103$~nm (dotted, green), $222$~nm (solid blue) and $458$~nm (dashed, orange), respectively. Here $X$ and $P$ are the mechanical position and momentum quadratures, respectively, in units of the interferometer readout range $\lambda_{\textsc{r}}/4=158.2$~nm (corresponding to the turning-points of the interferometer).}
\label{fig:Setup}
\end{figure}

For simplicity we first describe our scheme for two-port photon counting before generalizing to the multi-port case. The two-port case describes our experimental results and can be implemented with Mach-Zehnder-type interferometers. Figure~\ref{fig:Setup}(a) shows a schematic of our experiment. A weak coherent state is injected into the interferometer and interacts with a mechanical resonator in one of the interferometer arms. The two optical fields inside the interferometer then interfere on a beam-splitter and photon counting is performed on the two output ports. A single-photon click on one of the detectors gives an event-ready signal that the mechanics had interacted with the optical path-entangled number state $(\ket{10} + \ket{01})/\sqrt{2}$. If, instead, both detectors register a single photon then, due to second-order quantum interference, the optical state that interacted with the mechanical resonator within the interferometer must have been the 2-photon NOON-state $(\ket{20} - \ket{02})/\sqrt{2}$. In this case, the mechanical resonator was subject to a superposition of the identity operation and a two-photon radiation-pressure displacement, thus enhancing the size of the superposition by a factor of two compared with the single-photon detection case.

This operation can be described using the measurement- (or Kraus-) operator approach, which allows us to compute the mechanical state after the interaction and photon-counting measurement as well as the measurement outcome probabilities. Consider a mechanical resonator in the pure initial state $\ket{\psi_\textrm{in}}_\textsc{m}$ and coherent states $\ket{\alpha}_1 \ket{\alpha}_2$ in the interferometer arms 1 and 2, respectively (i.e.\ the state after the initial $50/50$ beamsplitter in Figure~\ref{fig:Setup}(a)). The state immediately after the interaction with mode 1 of the interferometer is then given by $e^{i \mu a_1^\dagger a_1 X_\textsc{m}} \ket{\alpha}_1 \ket{\alpha}_2 \ket{\psi_\textrm{in}}_\textsc{m}$, where $\mu$ is the single-photon optomechanical coupling strength, which quantifies the mechanical momentum transfer per photon in units of the mechanical quantum noise. For a cavity optomechanical system $\mu \propto g_0/\kappa = 4 \mathcal{F} x_0 / \lambda$, and is thus enhanced by the cavity finesse $\mathcal{F}$. For a single prompt reflection, $\mu = 4\pi x_0 / \lambda$. Here, $x_0=\sqrt{\hbar/m \omega}$ is the mechanical ground state size ($m$; effective mass, $\omega$; mechanical angular frequency), $X_\textsc{m} = x/x_0$ is the mechanical position operator in units of $x_0$, and $a_{1,2}$ are the annihilation operators for the interferometer arms 1 and 2. The two optical fields then interfere on a beamsplitter and photon counting is performed on the two output ports where $m$ and $n$ photons are detected in modes $1$ and $2$, respectively. The mechanical state after this interaction and measurement is $\ket{\psi_\textrm{out}}_{\textsc{m}} \propto {}_2\bra{n} {}_1\bra{m} B e^{i \mu a_1^\dagger a_1 X_\textsc{m}} \ket{\alpha}_1 \ket{\alpha}_2 \ket{\psi_\textrm{in}}_\textsc{m}$, where $B$ is the beam-splitter operator. Assuming a 50:50 beam-splitter, and including a static phase-shift $\phi$ in mode 2 of the interferometer, we obtain the measurement operator,
\begin{equation}
\Upsilon = \frac{e^{-|\alpha|^2}}{\sqrt{m!n!}}  \frac{\alpha^{m+n}}{(\sqrt{2})^{m+n}} (e^{i\mu X_\textsc{m}} + e^{i\phi})^m (e^{i\mu X_\textsc{m}} - e^{i\phi})^n \, .
\label{Eq:Upsilon}
\end{equation}
For the click event $\{m,n\}$ = $\{1,0\}$, this operation corresponds to a superposition of a mechanical displacement $e^{i\mu X_{\textsc{m}}}$ and the identity operation with a controllable phase $e^{i\phi}$. Using this measurement operator, the mechanical state after the operation is $\rho_\textsc{m}^\textrm{out} = \Upsilon \rho_\textsc{m}^\textrm{in} \Upsilon^\dagger / \mathcal{P}$, where $\rho_\textsc{m}^\textrm{in,out}$ are the input and output mechanical density matrices, respectively, and $\mathcal{P} = \integlim{X_\textsc{m}}{-\infty}{\infty} \Upsilon^\dagger \Upsilon \bra{X_\textsc{m}} \rho_\textsc{m}^\textrm{in} \ket{X_\textsc{m}}$ is the probability for obtaining the photon counting outcomes $m$ and $n$. See the Supplementary Material for further details of our theoretical model.

The mechanical position probability distribution of the state after the interaction and measurement is $\bra{X_\textsc{m}} \rho_\textsc{m}^\textrm{out} \ket{X_\textsc{m}} \propto \Upsilon^\dagger \Upsilon \bra{X_\textsc{m}} \rho_\textsc{m}^\textrm{in} \ket{X_\textsc{m}}$. The function $\Upsilon^\dagger \Upsilon$ is oscillatory in $X_{\textsc{m}}$, as obtaining a click in our interferometer gives information periodic in $X_{\textsc{m}}$, and can be interpreted as a filter acting on the initial mechanical position probability distribution. This oscillatory behaviour is intrinsically linked with the cubic nature of the full optomechanical radiation-pressure interaction $a_1^\dagger a_1 X_\textsc{m}$ and the non-linearity of photon counting. This allows our scheme to generate non-Gaussian states of motion, which is not possible in the more commonly considered linearized regime with quadratic interactions and linear measurements. It is also useful to note here that for pure states the position and momentum probability distributions are Fourier transforms of one another. Thus, the bimodal probability distribution generated by the superposition of identity and displacement operations gives rise to the interference fringes in the position quadrature.

\subsection{Experimental setup}

In our experiment, Figure~\ref{fig:Setup}(b), we use a high-stress $1.7\times 1.7$~mm Si$_3$N$_4$ membrane~\cite{Zwickl08} embedded in a $10\times 10$~mm Si-frame. The membrane has a thickness of $50\pm 2.5$~nm and, at our state-preparation-field wavelength of $795$~nm, has a measured reflectivity of $23.0\pm 0.5~\%$, while the frame has a reflectivity of $20.5\pm 0.2~\%$. The noise-power spectrum of the fundamental drum mode at $\omega_\textsc{m}/2\pi = 105.64 \pm 0.02$~kHz is shown in Figure~\ref{fig:Setup}(c). At room temperature and at atmospheric pressure the mechanical line-width (FWHM) was measured to be $\delta\omega_{\textsc{m}}/2\pi =  3.10 \pm 0.05$~kHz and the effective mass is on the order of $100$~ng, which comprises approximately $10^{16}$ atoms. In order to probe the regime where the optomechanical phase shifts are large, i.e. $\mu^2 \langle X_\textsc{m}^2 \rangle \gtrsim 1$, we use a ring-piezo to drive the membrane motion with random (thermal) fluctuations up to an RMS motion of 200~nm, see Supplementary Material for details.

The membrane is mounted in a way that allows optical access from both sides, and forms the central part of two folded Mach-Zehnder interferometers (MZI), see Figure~\ref{fig:Setup}(b). One interferometer is used for mechanical state preparation with photon counting, as illustrated in Figure~\ref{fig:Setup}(a), while the other is used for mechanical position readout using a balanced detector and a ${\sim}100~\mu$W laser at a wavelength of $632.8$~nm. For simplicity, both the state-preparation and the readout laser are operated continuous-wave for this proof-of-concept experiment. In the ideal case, they would both be pulsed, one after the other, however, as the state-preparation field is very weak, i.e. the probability of having a single photon present per mechanical period is low, an effective pulsed operation is implemented \emph{a posteriori} by the photon counters that have a timing resolution much smaller than a mechanical period. Our setup employs a compact polarization interferometer design that does not require active phase stabilisation~\cite{Ringbauer15}. The two arms of the MZI in Figure~\ref{fig:Setup}(a) are represented by orthogonal polarizations and the role of the beamsplitter is achieved by a half-wave plate (HWP), which allows for precise control of the splitting ratio. For the interaction with the membrane the two polarizations are separated using a calcite beam-displacer and recombined after reflection from the mechanical device with one polarization gaining a mechanical position dependent phase shift.

An APD click---either one of the detectors or a coincidence event within a $7.8$~ns window---triggers the balanced detector and recording of a $50~\mu$s long trace at a sampling rate of $100$~MS/s, see Figure~\ref{fig:Setup}(d). The mechanical quadratures $X$ and $P$, defined in units of the interferometer readout range ($\lambda_{\textsc{r}}/4$), are then extracted from a fit to this time trace. For small mechanical displacements the time trace is almost sinusoidal, but becomes overmodulated as the resonator displacement surpasses ${\sim} 100$~nm. In addition, for larger mechanical displacements we observe a mechanical-position-dependent amplitude modulation of the interferometer signal. We have taken the first order corrections due to this amplitude modulation into account when fitting the time traces, see Supplementary Material. For each type of click event we record ${\sim} 3000$ such time traces to create a phase-space histogram of the mechanical motional state (100 bins of width $0.034$ ($0.08$) in the weak (strong) drive regime, in units of the interferometer readout range). By using a combination of spectral and polarization filters the read-out beam transmitted through the membrane is suppressed below the dark-count level of our single-photon detectors of ${\sim}150$~Hz. The effect of these dark counts is a negligible decrease in fringe visibility, as discussed in the Supplementary Material.

\section{Results}
\subsection{Large phase-shift regime}

\begin{figure}[ht!]
\centering
\includegraphics[width=\linewidth]{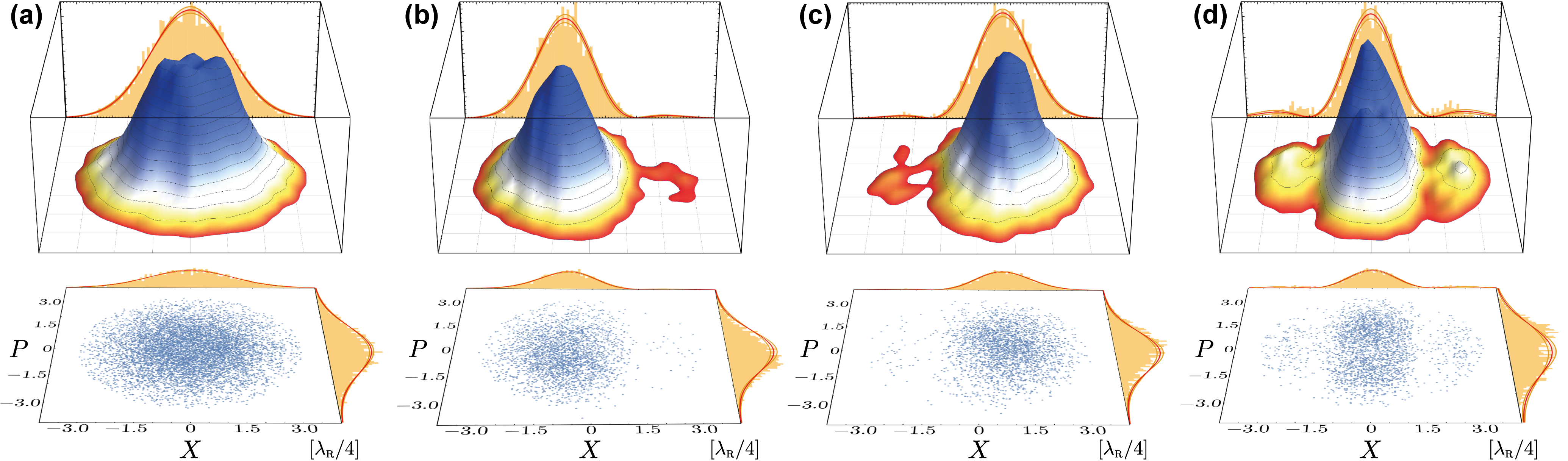}
\caption{\textbf{Observed mechanical states of motion showing interference fringes.}
Each plot shows the measured phase-space points (bottom), together with the corresponding mechanical position and momentum quadrature histograms, and a normalized 3D probability-density histogram of these points (top), with the mechanical position quadrature histogram reproduced again for easy visualization. The quadrature histograms contain a fit (solid, red line) using the theory model and corresponding 3$\sigma$ mean prediction bands (shaded orange). All axes are in units of the interferometer readout range (158.2~nm). (\textbf{a}) Measured initial Gaussian thermal state. (The truncation at the origin is due to finite resolution of our data acquisition.)
(\textbf{b}) Conditional motional state prepared via $\{m,n\}=\{0,1\}$ detection, which is shifted in phase-space along $-X$ and shows the start of a second peak in the right tail of the Gaussian envelope.
(\textbf{c}) Conditional state for $\{m,n\}=\{1,0\}$ detection, which is as B, but shifted along $+X$. Again, note the second peak, now on the left.
(\textbf{d}) Two photon ($\{m,n\}=\{1,1\}$) detection generates a mechanical fringe pattern in $X$ with twice the frequency of the single photon detection cases due to super-resolution of the measurement. Three maxima of the fringe pattern are observed. (As in (a), the finite resolution of our data acquisition leads to a truncation at the origin, which here manifests as a slightly double-peaked structure in the momentum distribution.)
In all cases the interferometer phase $\phi$ was set to $\pi/2$. Also note that the momentum quadrature remains close to the initial distribution for all of these measurements.}
\label{fig:Results}
\end{figure}

Figure~\ref{fig:Results} shows the measured mechanical phase-space distributions prepared via one- and two-photon measurements on a piezo-driven initial Gaussian thermal-state (Figure~\ref{fig:Results}(a)) with RMS position fluctuations of $198\pm 2$~nm. This corresponds to the regime of large optical phase-shifts, i.e. $\mu^2 \langle X_\textsc{m}^2 \rangle \gtrsim 1$. For single-photon detection, low frequency fringes are observed with a $\pi$ phase-shift between the detection events $\{m,n\}$ = $\{0,1\}$ (Figure~\ref{fig:Results}(b)) and $\{1,0\}$ (Figure~\ref{fig:Results}(c)). Moreover, we observe the start of the second fringe peak in the tails of the Gaussian envelope, which is on the right in Figure~\ref{fig:Results}(b) and the left in Figure~\ref{fig:Results}(c). In the case of a two-photon detection event $\{m,n\}$ = $\{1,1\}$ (Figure~\ref{fig:Results}(d)), the mechanical resonator interacted with an effective two-photon NOON state. Consequently, we observe phase super-resolution in the mechanical interference pattern at twice the fringe-frequency of the single photon cases. We would like to highlight that our scheme maps the fringe pattern onto the state of another bosonic mode and thus uses a time-reversed programme of quantum metrology for quantum-state-engineering applications. Thus, the super-resolution achieved here is a resource for state preparation. In a quantum regime, the fringe pattern observed can then be interpreted as either the quantum interference in the superposition state or as a consequence of the filter of the quantum measurement. This measurement-based technique provides a considerable advantage for ultimately generating non-classical states of mechanical motion and can be employed in other quantum optical systems. Note that all the states prepared (Figure~\ref{fig:Results}(b-d)) feature interference fringes in the position distribution, while the momentum quadrature remains close to the initial Gaussian distribution. The conditional mechanical states shown here were prepared with the phase set to $\phi=\pi/2$, which gives a fringe maximum in the centre of the distribution for the $\{1,1\}$ event.

We would like to additionally note here that our method can be utilised to determine the dimensionless optomechanical coupling strength $\mu$ by fitting to the fringe pattern observed. This technique requires a well calibrated position axis in units of the mechanical ground state size and can be performed for any mechanical thermal occupation.

\subsection{Small phase-shift regime}

\begin{figure}[!h]
\begin{center}
\includegraphics[width=0.6\linewidth]{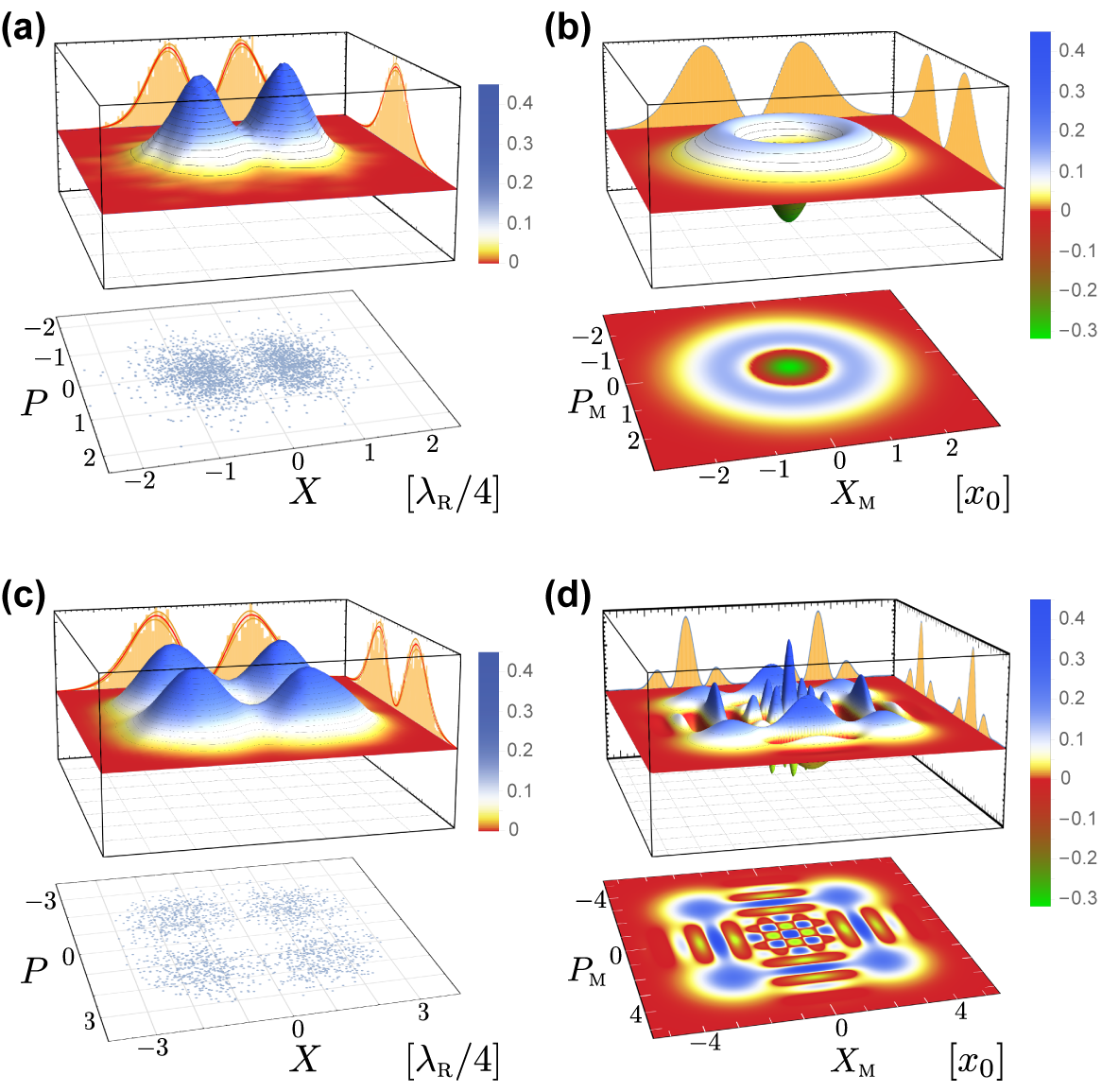}
\vspace{-0.7em}
\caption{\textbf{Non-Gaussian mechanical states prepared with weak and strong single-photon coupling.}
(\textbf{a}) Experimentally observed mechanical phase-space distribution for the detection event $\{m,n\}=\{1,1\}$ with $\phi=0$ in the weak coupling and weak drive regime, in units of the interferometer readout range. The position probability distribution here has the same form as that of a single phonon Fock state. 
(\textbf{b}) Theoretical Wigner function of a single phonon Fock state, plotted in units of the mechanical quantum noise. This state would be obtained when applying our scheme with $\{m,n\}=\{1,1\}$ with $\phi=0$ as in panel~(a), to the mechanical ground state with $\bar{n} \sim 0$ and for arbitrarily small $\mu$. Note that the Fock state has a non-Gaussian momentum distribution whereas our data has a Gaussian momentum distribution due to the thermal initial state.
(\textbf{c}) Experimentally observed mechanical phase-space distribution for a sequence of two detection events $\{m,n\}=\{1,0\}$ with $\phi=\pi/2$, separated by a quarter period of mechanical motion, in the weak coupling and weak drive regime. Note that both the position and momentum probability distributions have the same form as those of a single phonon Fock state.
(\textbf{d}) Theoretical Wigner function expected from applying the same preparation scheme as in panel~(c) to a mechanical resonator in the ground state ($\bar{n}\sim 0$) and with a coupling strength of $\mu = 5$. This state is known as a `quantum compass state'.
}
\label{fig:SmallMu}
\end{center}
\end{figure}

Our scheme can also generate non-Gaussian states of motion in the regime of small optomechanical phase shifts, i.e.\ $\mu^2 \langle X_\textsc{m}^2 \rangle \ll 1$. Indeed, mechanical non-classicality, in the form of Wigner negativity, can be generated independent of the coupling strength, providing a promising route to generate and explore macroscopic mechanical quantum states even for weak single-photon optomechanical coupling. For small $\mu$, applying our scheme to the mechanical ground state for $\{m,n\}=\{0,1\}$ with $\phi=0$ yields $(e^{i\mu X_\textsc{m}} - 1)\ket{0} \simeq (i\mu X_\textsc{m})\ket{0} = i\mu 2^{-1/2}\ket{1}$. Here, $\ket{1}$ denotes a single phonon Fock state and the mechanical position operator in terms of the phonon annihilation ($b$) and creation ($b^\dagger$) operators is $X_{\textsc{m}} = 2^{-1/2} (b+b^\dagger)$. This state is a result of quantum interference and cannot be described classically~\cite{Zavatta07, Kleckner08, Pepper12}. Note that in this regime the detection event $\{m,n\}=\{1,1\}$ generates a similar state. The Wigner function of this mechanical single-phonon Fock state and its two conjugate quadrature distributions are shown in Figure~\ref{fig:SmallMu}(b). A mechanical Fock state has a rotationally invariant distribution in phase-space and has the quadrature distribution $\textrm{Pr}(X_\textsc{m}) = 2 \pi^{-1/2} X_\textsc{m}^2 e^{-X_\textsc{m}^2}$ for all quadrature angles. For $\mu^2 \langle X_\textsc{m}^2 \rangle \ll 1$ the filter for these detection events is $\Upsilon^\dagger \Upsilon \propto \mu^2 X_{\textsc{m}}^2$. Thus, we note that the position probability distribution after this operation has the same functional form as a single-phonon Fock state for any thermal occupation, not just the ground state. More specifically, the position probability distribution is a product of a Gaussian with $X_\textsc{m}^2$ in both cases, where the initial thermal occupation sets the overall width of the Gaussian envelope of the distribution. This comparison identifies that there is no qualitative transition in the position probability distribution between application of our scheme in the large thermal noise regime and application on the mechanical ground state. Thus, careful calibration and full tomography becomes more important in order to certify mechanical non-classical behaviour. These points will be discussed further in the final section of this manuscript.

To observe this type of fringe pattern we use the detection event $\{m,n\}=\{1,1\}$ on a mechanical thermal state with RMS position fluctuations of $91\pm1$~nm. The observed mechanical phase-space distribution and quadrature distributions are shown in Figure~\ref{fig:SmallMu}(a). The measured position probability distribution is in good agreement with the theoretical prediction. Note that the momentum distribution remains Gaussian due to the thermal nature of the initial state. 

The approach presented here has the additional versatility that multiple measurements in time can be used to generate more complex states of mechanical motion. As an example, Figure~\ref{fig:SmallMu}(c) shows the state resulting from an experimental sequence of two $\{m,n\}=\{1,0\}$ events, separated by a quarter period of mechanical harmonic evolution. Notably, this state reproduces the qualitative form of the marginal probability distribution of the Fock state not only in position, but also in momentum, thus highlighting the importance of full state reconstruction. When applied to a low temperature state this procedure would generate quantum compass states~\cite{Zurek01}, which are superpositions of four coherent states at cardinal positions in phase space, see Figure~\ref{fig:SmallMu}(d). These highly non-classical states have very sharp phase-space features and are of considerable interest for quantum information and quantum sensing applications.

\subsection{Generalisation to larger NOON states}

\begin{figure}[!h]
\begin{center}
\includegraphics[width=1\linewidth]{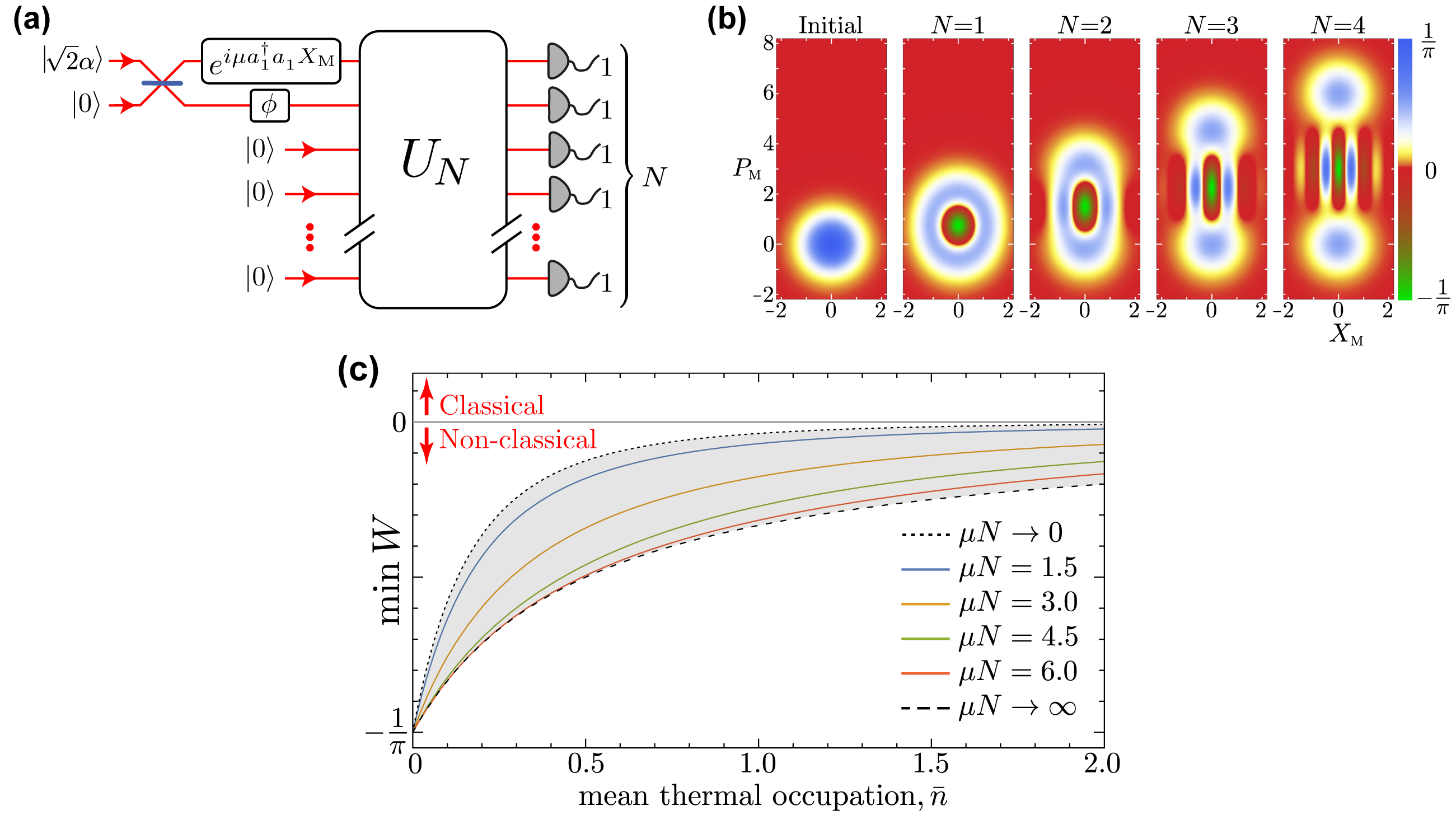}
\vspace{-2em}
\caption{\textbf{Generation of larger superposition states.}
(\textbf{a}) Optical setup requiring two input coherent states, a linear optical network $U_N$, and single photon counters. One of the two input coherent states interacts with the mechanical resonator and then the fields are projected onto an optical NOON state, which generates a mechanical superposition state with a separation that increases with $N$, see text.
(\textbf{b}) Simulated Wigner functions for an initial mechanical ground state ($\bar n =0$) and superposition states prepared via multi-photon detection. From left to right: initial ground state, and mechanical states prepared using 1-, 2-, 3-, 4-photon NOON-state projections for $\mu=1.5$ and $\phi = 0$. Note that these mechanical states are generated via a superposition of the identity operation and a displacement and are thus not symmetric around the origin. 
(\textbf{c}) Minimum of the Wigner distribution ($\min W$) as a function of the initial thermal occupation $\bar{n}$ for various values of $\mu N$. Note that our scheme generates Wigner-negativity even in the limit of weak coupling $\mu N\to 0$ (dotted, black line) and saturates for large coupling $\mu N\to \infty$ (dashed, black line). All experimental configurations (i.e.\ arbitrary values of $\mu, N$ and $\bar n$) result in states within the grey shaded area, and always feature Wigner-negativity. Moreover, for arbitrary single-photon coupling strength $\mu$ and arbitrary NOON-state size $N$ the generated states achieve the maximum possible negativity of $-1/\pi$, as $\bar n \to 0$. The coloured solid lines correspond to $N=$ 1, 2, 3, 4 multi-photon coincidence events, for $\mu=1.5$, as in panel (b).}
\label{fig:Multiport}
\end{center}
\end{figure}

Our scheme has the advantage that it can be easily extended to generate mechanical superposition states with increasing separation size by  changing only the optical heralding measurement. This extension requires two coherent states, one of which interacts with the mechanical system, while the other acquires a static phase shift, together with $N-2$ vacuum ancilla modes injected into an optical $N$-port interferometer, see Figure~\ref{fig:Multiport}(a). $N$-fold single-photon coincidence detection at the output of the $N$-port then provides an event-ready signal that the mechanical resonator interacted with an optical NOON-state. The mechanical resonator is thus subject to a superposition of a radiation-pressure interaction with $0$ or $N$ photons which enhances the separation in the superposition state, see Figure~\ref{fig:Multiport}(b). Here, the high-frequency phase-superresolving fringes of the optical NOON-state are mapped onto the mechanical position probability distribution. In a cavity optomechanics experiment, one would use a number of photon counters $N$ that provide a practical experimental duration given the heralding probability and repetition rate of the experiment. The heralding probability for this operation including projection onto NOON states is $\mathcal{P}_N(\bar{n}) = \frac{2}{N^N} e^{-2|\alpha|^2} |\alpha|^{2N} ( 1 - (-1)^N \exp[-{\textstyle \frac{1}{4}}\mu^2(1+2\bar{n})N^2] \cos(N \phi) )$. See the Supplementary Material for further details and a mathematical description for this NOON state projection technique.

Importantly, the states generated by our scheme can exhibit strong negativity of the Wigner quasi-probability distribution independent of the optomechanical coupling strength $\mu$. As shown in Figure~\ref{fig:Multiport}(c), this negativity approaches zero from below asymptotically with increasing $\bar{n}$, and thus the scheme is resilient against finite initial thermal occupation. Furthermore, we would like to highlight that for large $\mu N$ this negativity scales with $(1+2\bar{n})^{-1}$. To the best of our knowledge, this is the most resilient scaling of Wigner negativity with $\bar{n}$ reported thus far. Operations generating Wigner negativity when applied to thermal states have been reported previously, and the scaling found here should be compared to the previous best result of single-quanta addition~\cite{Zavatta07}, which was applied in the quantum optics community for single photon addition to an optical thermal state. In that work the Wigner negativity generated goes with $(1+2\bar{n})^{-2}$. Though this latter scheme has the advantage of operating in the weak coupling regime, our work demonstrates that more resilient scaling is achievable for large $\mu N$. Please refer to the Supplementary Material for more quantitative details. Moreover, since low amplitude optical coherent states are used, our scheme is robust against optical loss and inefficiency as the single-photon event-ready signals selects the cases where no light was lost from the system.

\section{Discussion and Conclusion}

We have introduced a technique that exploits the nonlinearity of multi-photon counting to provide a more feasible path to generate momentum superposition states of motion of a macroscopic mechanical resonator. Additionally, we have performed a proof-of-concept experimental demonstration of this technique, which made the first observation of mechanical interference fringes within a thermal distribution. These fringes, albeit at a classical level, are an exciting step towards generating a quantum superposition state by applying our technique to a low entropy initial state. Additionally, there is no qualitative or quantitative difference between the quantum and classical predictions for the fringes in the mechanical position probability distribution for all values of $\mu$ and $\bar{n}$, which highlights that the appearance of such fringes is not a sufficient condition for non-classicality. This point further motivates the importance of precise calibration of the position quadrature measurements, characterization of the initial state, and full quantum state reconstruction to be able to claim the observation of non-classical mechanical motion. While our continuous-wave readout technique is suitable for the regime experimentally tested here, this technique cannot resolve displacements below the size of the ground state, which is required for mechanical quantum state reconstruction. Different techniques, such as quantum non-demolition pulsed quadrature measurements~\cite{Vanner2011,Vanner2013}, or two-toned driving schemes~\cite{Lei2016}, can allow this limit to be surpassed~\cite{Vanner15}. A detailed discussion of position measurements and mechanical state reconstruction is beyond the scope of the present work and the reader is encouraged to refer to these references for further details.

Looking ahead, a promising experimental approach to generating significant non-classicality with our scheme is to use MHz-frequency mechanical resonators in high bandwidth optical cavities that operate well outside the resolved-sideband regime. Rapid experimental progress is currently being made in this parameter regime with prominent examples of large single-photon coupling in solid-state systems including Refs~\cite{Wilson15, Krause15, Leijssen17}. For this configuration, measurement and feedback is well suited to pre-cooling the mechanical motion to low initial occupations before applying our non-Gaussian operation described here. Indeed, recent experimental work~\cite{Wilson15, Krause15} has demonstrated large optomechanical coupling for precision position measurements and feedback cooling to near the ground state. Moreover, as detailed therein, near-future improvements will allow cooling to thermal occupations well below unity, even from room temperature. Utilising the parameters of Ref.~\cite{Wilson15} (mechanical frequency $\omega_\textsc{m}/2\pi = 4.3$~MHz; cavity decay rate $\kappa/2\pi = 440$~MHz; vacuum optomechanical coupling rate $g_0/2\pi = 20$~kHz; thermal occupation achieved $\bar{n} = 5.3$), the coupling strength achievable is $\mu = 5\times10^{-5}$, and the Wigner negativity generated by a single photon click is $2 \times 10^{-4}$. Excitingly, employing our protocol to near-future improvements expected with this system that provide further cooling to a thermal occupation of $\bar{n}=0.1$~\cite{Wilson15}, the Wigner negativity generated is $-0.18$. See Figure~4 for a plot of the Wigner negativity with thermal occupation and the Supplementary Material for quantitative details. We would also like to highlight that ultracold atom optomechanical systems~\cite{Brennecke2008} can achieve $\mu > 1$ providing a present-day system that can generate large quantum superpositions with our scheme. Furthermore, it has been proposed that an array of mechanical devices within a cavity can lead to large single-photon optomechanical coupling via the collective enhancement~\cite{Xuereb2012}. Assuming a thermal occupation of $\bar{n}=0.1$ for such approaches that operate with large single-photon coupling, the Wigner negativity that can be generated is $-0.27$. This large negativity is approaching the maximum value achievable of $-\pi^{-1} \simeq -0.32$. Moreover, to illustrate the resilience of our scheme with $\bar{n}$, the Wigner negativity achievable for a mechanical system with an initial occupation of $\bar{n}=10$ for large $\mu N$ is $-0.02$.


Our scheme offers five main advantages: (i) a path is provided to generate mechanical momentum superposition states by projecting an optical pulse onto a NOON state. This scheme does not require non-classical optical input states that are difficult to prepare for narrow-bandwidth cavity interactions. Larger mechanical superposition states can be prepared by changing only the optical measurement and projecting onto larger NOON states. (ii) Our process generates non-classical mechanical states---signified by negativity in the Wigner quasiprobability distribution---without the need for strong single-photon coupling and (iii) with more favourable scaling against the initial thermal occupation than previously reported. (iv) The protocol can be performed quickly compared to the mechanical period, as it operates outside the resolved-sideband regime, making it much easier to complete the experiment before thermal decoherence sets in. (v) Our scheme is resilient against optical loss, as photon counting with low amplitude input coherent states can be used. This combination of advantages thus dramatically improves the feasibility to generate and observe mechanical momentum superposition states---a key outstanding goal of the field---and the high-visibility mechanical fringes observed here are a key step towards achieving this goal.

\ack
We would like to thank C.~M.~Caves, M.~S.~Kim, G.~J.~Milburn, I.~Pikovski, J.~Schmiedmayer, and F.~Shahandeh for helpful discussions and B.~Duffus and T.~Vulpecula for experimental assistance. This work was supported by an ARC Discovery Project (DP140101638), the ARC Centres for Engineered Quantum Systems (CE110001013) and Quantum Computation and Communication Technology (CE110001027), and the Engineering and Physical Sciences Research Council (EP/N014995/1). AGW acknowledges support through a UQ Vice-Chancellor's Research and Teaching Fellowship.

\section*{References}

\providecommand{\newblock}{}

\bibliographystyle{iopart-num}


\end{document}